\documentclass[a4paper]{jpconf}
\usepackage{graphicx}
\graphicspath{{figures/}}

\begin{document}
\title{Measurement of scintillation and ionization yield with high-pressure gaseous mixtures of
Xe and TMA for improved neutrinoless double beta decay and dark matter searches}

\author{Y~Nakajima, A~Goldschmidt, H~S~Matis, D~Nygren, C~Oliveira, J~Renner}

\address{Lawrence Berkeley National Laboratory, Berkeley, CA 94720, USA}

\ead{YNakajima@lbl.gov}


\begin{abstract}
Liquid Xe TPCs are among the most popular choices for double beta
decay and WIMP dark matter searches. Gaseous Xe has intrinsic
advantages when compared to Liquid Xe, specifically, tracking capability and better energy
resolution for double beta decay searches. The performance of gaseous
Xe can be further improved by molecular additives such as
trimethylamine(TMA), which are expected to  (1) cool down the
ionization electrons, (2) convert Xe excitation energy to TMA ionizations through
 Penning transfer, and (3) produce scintillation and
 electroluminescence light in a
more easily detectable wavelength (300 nm).
These features may provide better tracking and energy resolution for
double-beta decay searches. They are also expected to enhance columnar
recombination for nuclear recoils, which can be used for searches for
WIMP dark matter with directional sensitivity.
We constructed a test ionization chamber
and  successfully measured scintillation and ionization yields at high
precision with various Xe and
TMA mixtures and pressures. 
We observed 
the Penning effect and an increase in recombination
with the addition of TMA. However, many undesired features
for dark matter searches, such as  
strong suppression of the scintillation light and no sign of recombination
light, were also found. 
This work has been carried out within the context of the NEXT collaboration.
\end{abstract}

\section{Introduction}

Xenon TPCs are among the most popular and sensitive detector
technologies to search for WIMP dark matter and neutrinoless double
beta decay. 
While liquid Xe is more commonly used for such applications,
gaseous Xe
has intrinsic advantages over liquid Xe. For neutrinoless double beta
decay search, gaseous Xe provides superior energy resolution and
tracking capability. Those excellent capabilities are demonstrated in
Refs.~\cite{Alvarez:2012kua,Alvarez:2013gxa} and being carried out within the NEXT experiment~\cite{Granena:2009it}.

In addition, there is a potential capability of performing dark matter
searches with directional sensitivity utilizing columnar
recombination~\cite{Nygren:2013nda,Yasu-simulation}.  
To put this idea into practice, it is
essential to efficiently cool ionized electrons. The
directional sensitivity can be further enhanced by the Penning effect,
which converts primary excitation into an ionization signal. 
Hence, doping with molecular additives would be essential to achieve 
enhancement of columnar recombination.

We are exploring possible performance improvements by adding
trimethylamine(TMA) to gaseous Xe. We report measurements of
ionization and scintillation yields with a test ionization
chamber that we constructed at Lawrence Berkeley National Laboratory.

\section{Performance improvements with TMA}

For typical Xe TPCs, observable signals are scintillation light from
excited Xe ($\textrm{Xe} \to
\textrm{Xe}^*, \textrm{Xe}^*+\textrm{Xe} \to \textrm{Xe}^*_2 \to 2\textrm{Xe}(g.s.) + h\nu$) and ionized electrons ($\textrm{Xe} \to
\textrm{Xe}^+ + e^-$). 
The relative size of those signals can be altered with the presence of sufficient
recombination, $\textrm{Xe}^+ + e^- \to \textrm{Xe}^*$ and 
$ \textrm{Xe}^*+\textrm{Xe} \to \textrm{Xe}^*_2 \to
2\textrm{Xe}(g.s.) + h\nu$, which effectively converts ionization signals
into the scintillation signals. 
A sufficient amount of columnar recombination would be required for
a dark matter search with directional sensitivity. However, due to large
electron diffusion, such recombination in a pure Xe medium is expected
to be negligible. 

TMA is known to have a large inelastic cross section for electrons
due to various vibrational and rotational modes, and provides efficient
cooling of ionization electrons, which is essential for the recombination processes.
In addition, the ionization potential of TMA (7.85~$\pm$~0.05~eV~\cite{NIST:TMA}) is well matched to
the energy level of the first excited state of Xe (8.3~eV~\cite{NIST:Xe}). This allows a
Penning transfer ($\textrm{Xe}^* + \textrm{TMA}\to \textrm{Xe} + \textrm{TMA}^+
+ e^-$), which increases the intrinsic amount of ionization that would
undergo columnar recombination. 
We also expect that the charge exchange process, 
$\textrm{Xe}^+ + \textrm{TMA}\to \textrm{Xe} + \textrm{TMA}^+$,
would occur quickly. With efficient Penning and
charge-exchange transfers, initial excited and ionized Xe would be
largely converted into ionized $\textrm{TMA}^+$ and electron pairs.
Finally, TMA produces fluorescence light at $\sim 300$~nm,
which is much more easily detectable compared to the light from Xe
($\sim 170$~nm) because of typically higher PMT quantum efficiency at 300~nm.
If recombination of ionized TMA produces similar
fluorescence light, via $\textrm{TMA}^+ + e^- \to \textrm{TMA}^* \to
\textrm{TMA}(g.s.) + h\nu (300~\textrm{nm})$, then the detection
efficiency for columnar recombination would be significantly increased.

In addition, Penning transfer increases the number of ionized electrons,
which improves statistical precision of the energy 
measurement and track reconstruction.
Tracking performance would also be improved with TMA by reducing
the diffusion of ionization electrons~\cite{Gonzalez-Diaz:2015oba}.
These features would further benefit neutrinoless double-beta decay searches
in gaseous Xe.

\section{Experimental setup}

\begin{figure}[h]
\begin{center}
\includegraphics[width=0.8\columnwidth]{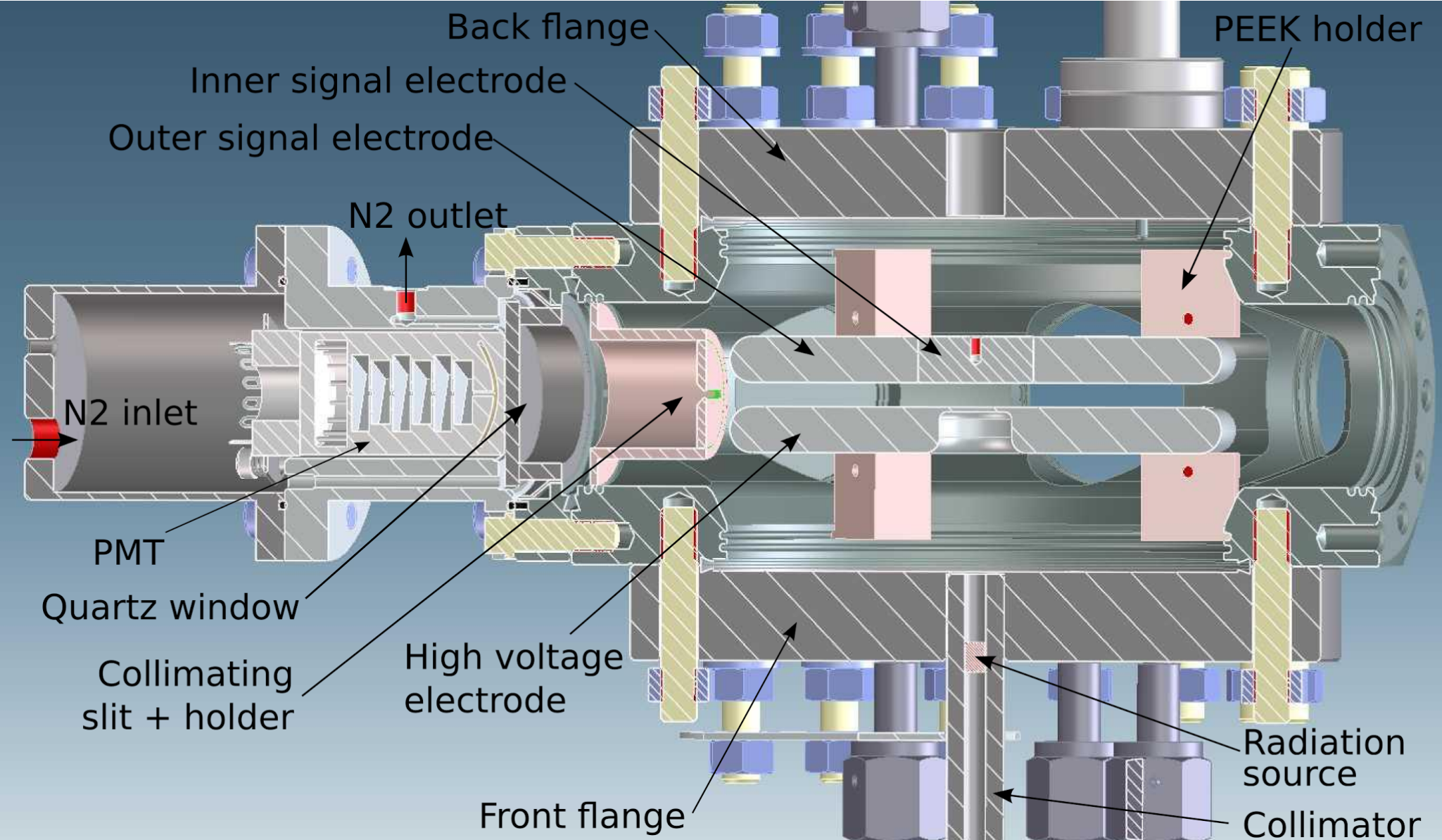}
\end{center}
\caption{\label{fig:teapot_drawing}Schematic of the ionization chamber.}
\end{figure}

In order to test the properties of the gas mixtures of Xe and TMA, especially 
Penning transfer efficiency and fluorescence light from recombination
processes, we built a test ionization chamber as shown in Fig.~\ref{fig:teapot_drawing}.

A pair of parallel plate electrodes with gap width of 5~mm is placed in the
chamber that can hold pressurized gas up to 8~bar. 
One of the electrodes is separated into a 2.5 cm-diameter inner electrode and
an outer electrode. The signal induced in the inner electrode is used for the
analysis, in order to avoid the effects from the charge produced
outside the gap between the electrodes.
Four PMTs are placed at the circumference of the electrodes to detect
scintillation light from the gas between the electrodes. A long-pass filter
that blocks light below 250~nm is installed in front of one of the
PMTs, so that this PMT is only sensitive to the scintillation light
from TMA, eliminating the Xe VUV light.
We irradiated the gas between the electrodes with $\sim$60~keV gamma
rays from a 10~mCi ${}^{241}$Am radioactive source. 
Currents from electrodes and PMTs were read with pico-ammeters in DC
mode.
A more detailed description of the setup can be found in
Ref.~\cite{Oliveira2015742}.

\begin{figure}[h]
\begin{center}
\includegraphics[width=0.8\columnwidth]{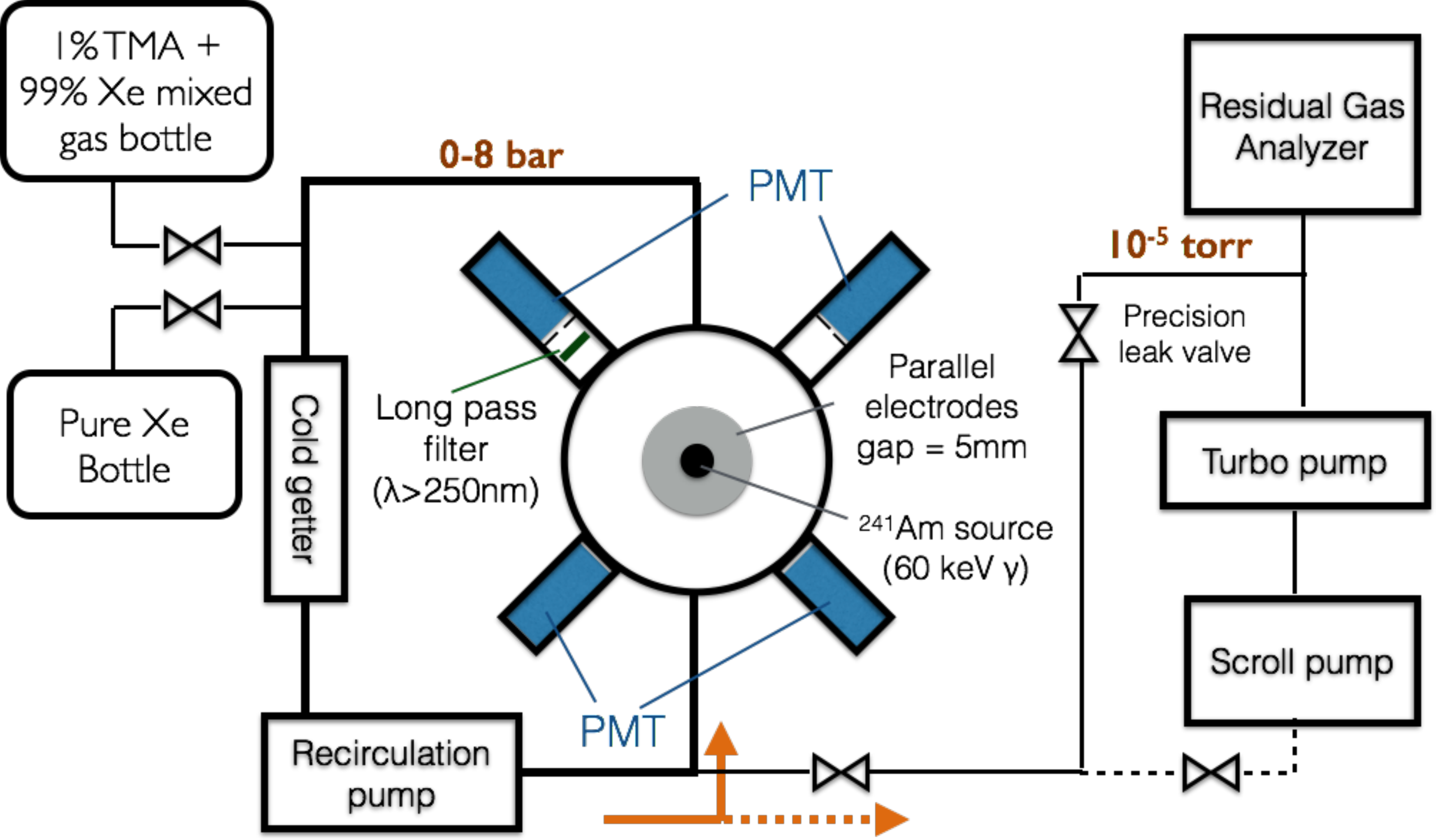}
\end{center}
\caption{\label{fig:teapot_diagram}Schematic diagram of the gas flow
  system.}
\end{figure}

\begin{figure}[h]
\includegraphics[width=24pc]{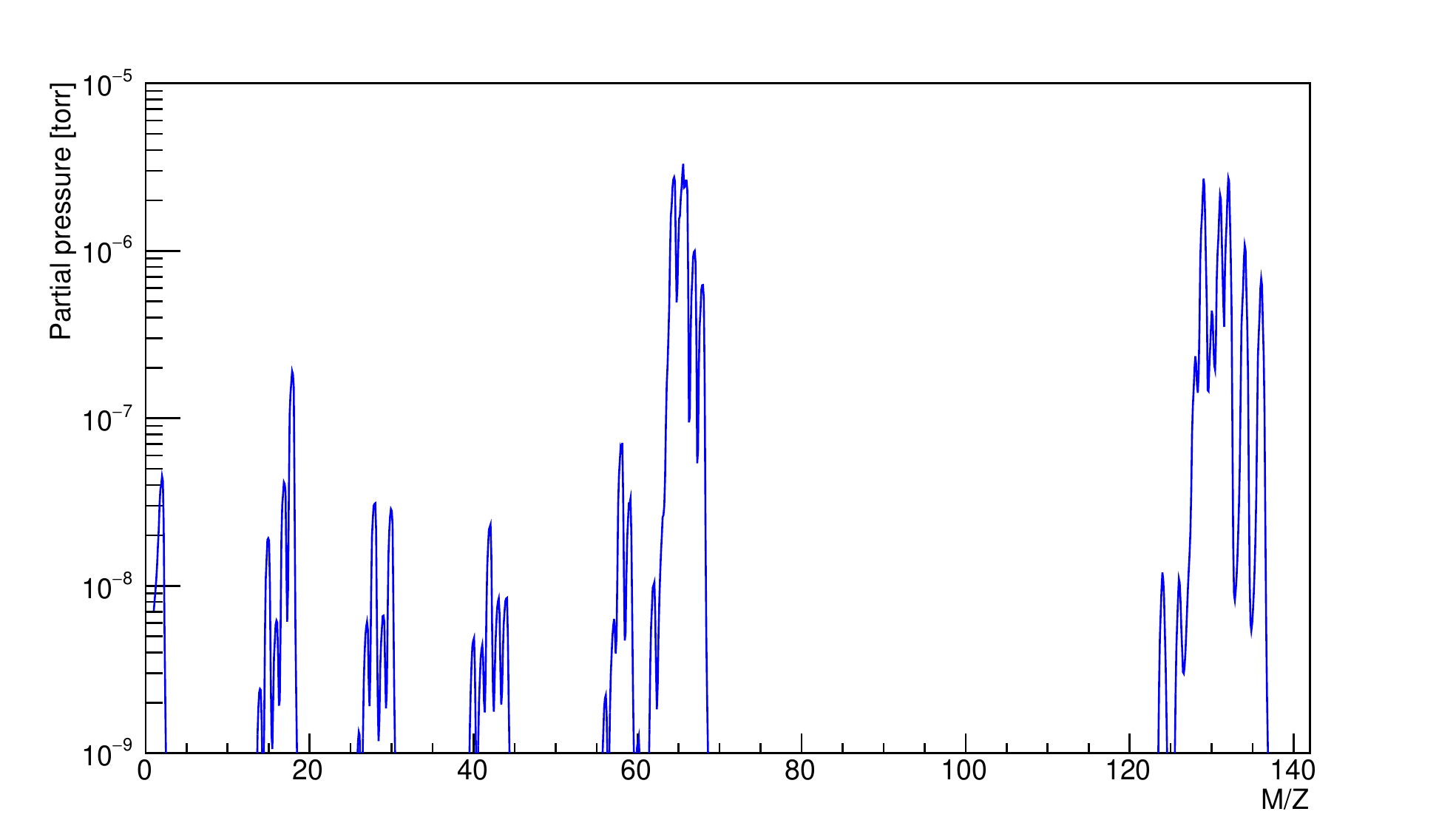}\hspace{0pc}%
\begin{minipage}[b]{14pc}\caption{\label{fig:RGA}Example output of
    an RGA analog mass scan for a  Xe+TMA gas mixture. The horizontal
    axis presents the 
    mass divided by the unit charge. Peaks around $M/Z = 130 (65)$
    correspond to singly(doubly) ionized Xe atoms. Peaks around
    $M/Z$ of 58, 42 and 30 are mostly from TMA and its
    fragments. Residual components of H${}_2$(M/Z=2), H${}_2$O(M/Z
    =18) and N${}_2$(M/Z=28) are also shown.}
\end{minipage}
\end{figure}

The gas flow in our system is illustrated in
Fig.~\ref{fig:teapot_diagram}.
Once the chamber is filled with the gas mixture, it is sealed
and the gas inside is purified by continuous circulation through a SAES MC50-702F
room temperature gas filter (``Cold getter'' in Fig.~\ref{fig:teapot_diagram}).
While this filter is essential for removing oxygen, which blocks
the scintillation light from Xe, we found that the filter also takes a significant amount of TMA from the Xe +
TMA mixture. We noticed that the TMA concentration in the gas mixture decreases shortly after introducing the gas into the system, and then
stabilizes after about one hour of recirculation.
This makes it difficult to precisely control the TMA fraction in the
chamber, and hence we
continuously monitored the gas composition using a residual gas
analyzer (RGA).
As shown in Fig.~\ref{fig:teapot_diagram}, a small fraction of the gas inside is continuously introduced into the RGA through the
precision leak valve. The gas pressure around the RGA was adjusted
to $\sim 10^{-5}$~torr.
Figure~\ref{fig:RGA} shows a typical mass spectrum of the Xe+TMA mixture, obtained
with the RGA. Singly and doubly ionized Xe, as well as TMA mass peaks,
can be seen. We monitored the relative height of Xe and TMA pressures,
and extracted the TMA fraction
in the main chamber. The absolute calibration of the TMA fraction was done
by introducing 1\% TMA and 99\% Xe mixture to the RGA  directly, without
the filter.

\section{Results with pure Xe}

\begin{figure}[h]
\begin{center}
\includegraphics[width=18pc]{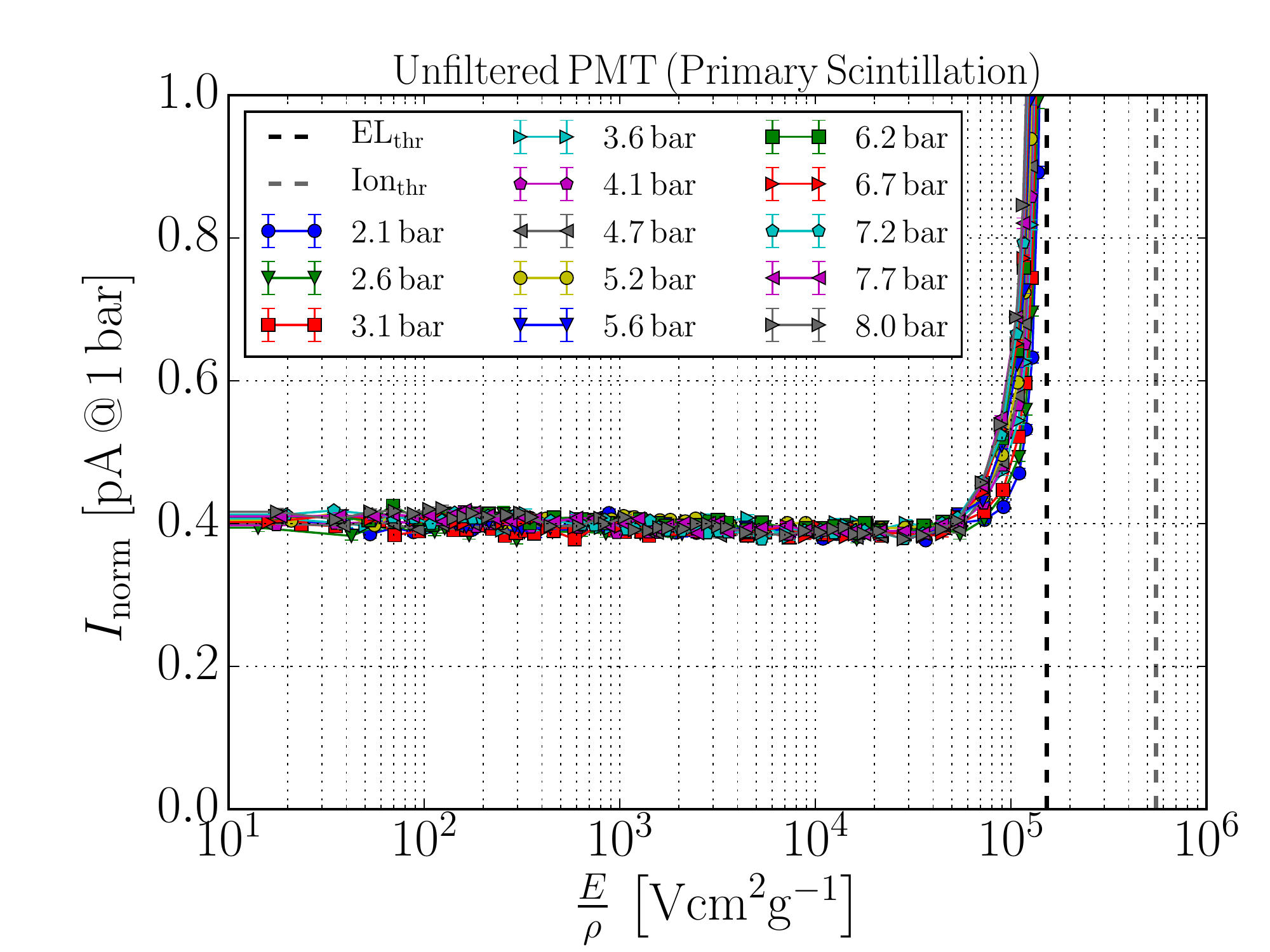}
\includegraphics[width=18pc]{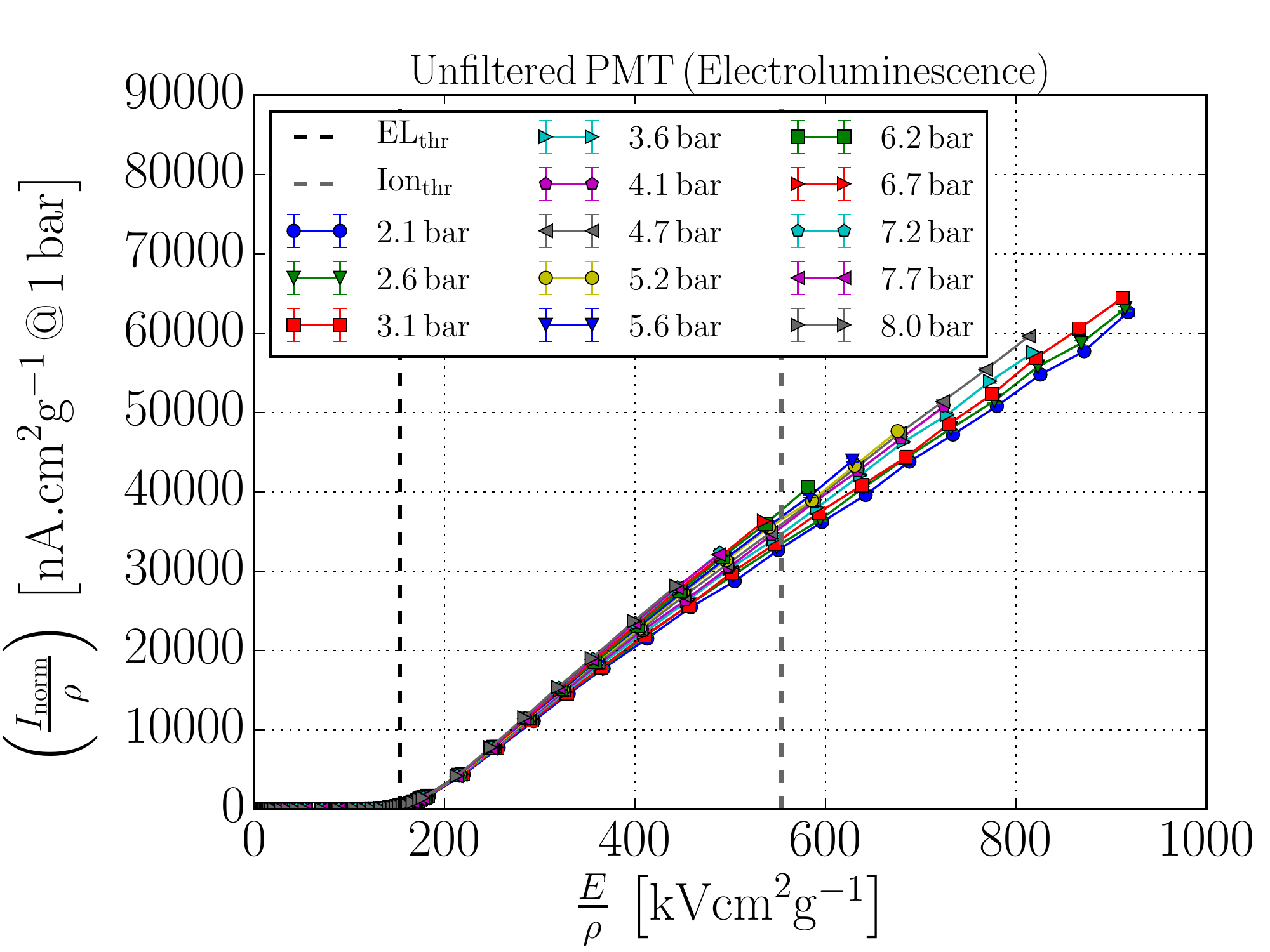}
\end{center}
\caption{\label{fig:purexe_pmt4} Light yield of pure Xe gas as a
  function of external electric field for different
  total pressures. (Left) Results in the primary scintillation
  region. (Right) Results in the Electroluminescence region.}
\end{figure}

\begin{figure}[h]
\includegraphics[width=18pc]{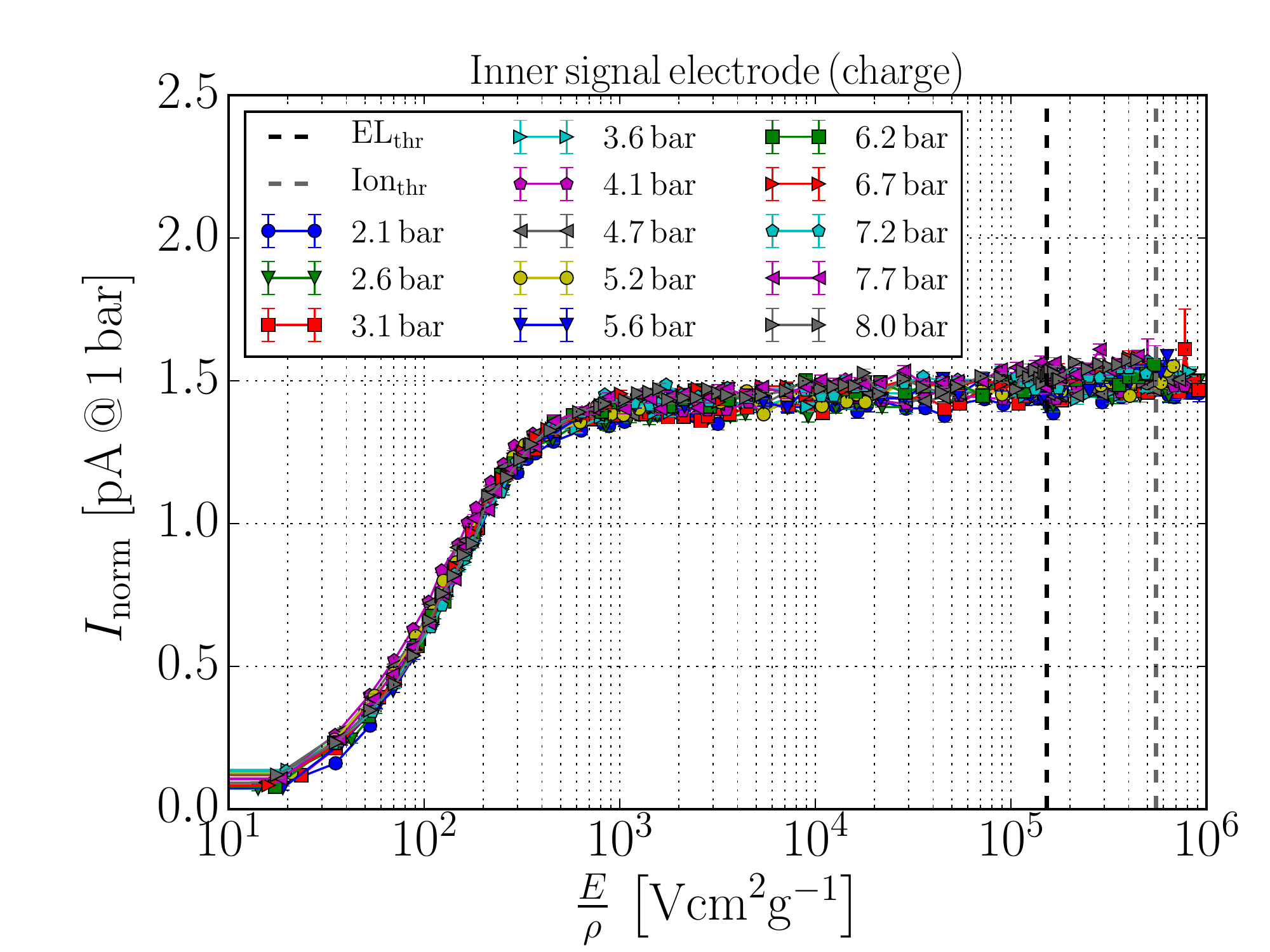}\hspace{2pc}%
\begin{minipage}[b]{18pc}\caption{\label{fig:pureXe_innerElectrode}Charge
  yield of pure Xe gas as a function of external electric field for different total pressures.}
\end{minipage}
\end{figure}

The system was initially tested with pure Xe gas~\cite{Oliveira2015742}.
Figure~\ref{fig:purexe_pmt4} shows results of light yield measurements
at various electric fields and pressures. The amount of energy
deposition in the active region, which changes with pressure, is corrected using a GEANT4-based
Monte Carlo simulation. Primary scintillation light for $E/\rho < 10^5~[\textrm{Vcm}^2\textrm{g}^{-1}]$
and proportional amplification at higher electric fields
(electroluminescence light) are clearly
seen. The results are consistent among the different pressures up to 8~bar.
Figure~\ref{fig:pureXe_innerElectrode} shows the current induced in
the electrode due to ionization processes. Again, good agreement between
different pressures was seen. The charge yield is relatively flat for
$E/\rho > 10^3~[\textrm{Vcm}^2\textrm{g}^{-1}]$, which  indicates an accurate
charge correction efficiency at various conditions. In addition, this flatness 
which extends to higher fields of $E/\rho \approx
10^5~[\textrm{Vcm}^2\textrm{g}^{-1}]$ suggests
that the increase in scintillation yield at the same higher field
region is due to proportional
amplification without charge amplification (avalanche).
The reduction of the charge yield at the low field region of
$E/\rho < 10^3~[\textrm{Vcm}^2\textrm{g}^{-1}]$ is likely due to
electrons diffusing out.
Those results are in good agreement with the previous measurements~\cite{Monteiro:2007vz,Freitas:2010zza}.
A more detailed description of test results with pure Xe can be found
in Ref.~\cite{Oliveira2015742}.

\section{Results with Xe and TMA mixture}

\subsection{Charge yield}
\begin{figure}[h]
\includegraphics[width=22pc]{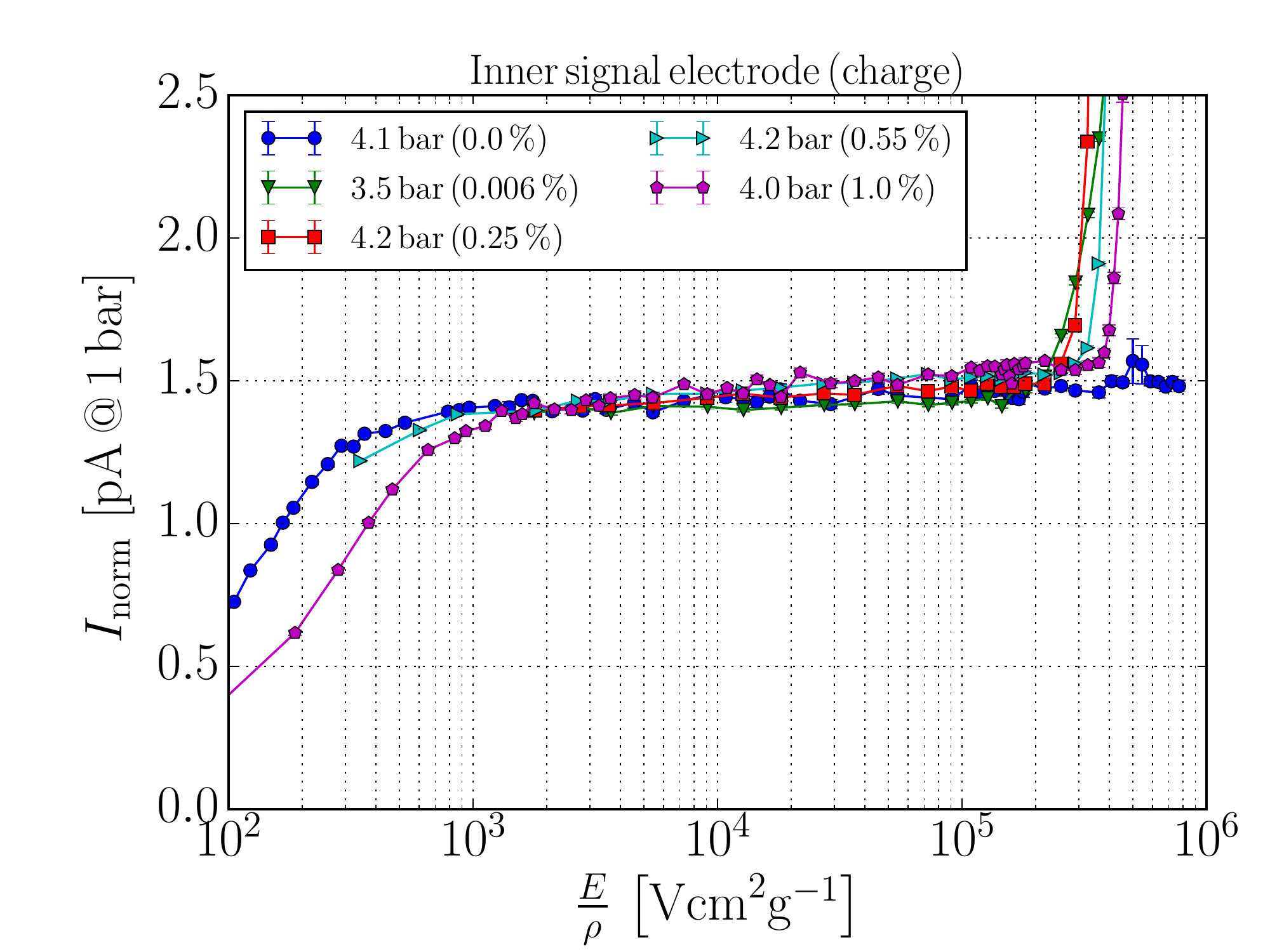}\hspace{1pc}%
\begin{minipage}[b]{14pc}
\caption{\label{fig:XeTMA-4bar-charge}Charge yield of gas mixture of
  TMA and  Xe as a function of external electric field for a total
  pressure of approximately 4~bar and various TMA concentrations.}
\end{minipage}
\end{figure}

Figure~\ref{fig:XeTMA-4bar-charge} shows the charge yield observed at the inner electrode at
approximately 4 bar total pressure and for various TMA concentrations. 
We found that the slope of charge yield vs. electric field at 
$10^3 < E/\rho < 10^5~(\textrm{Vcm}^2\textrm{g}^{-1})$
increases as the TMA fraction increases. That can be interpreted as an
effect of increased recombination, since recombination is larger at lower electric field and becomes
negligible at high electric field. Therefore, the larger slope
indicates larger amount of recombination at a low electric field with
more TMA. 
Furthermore, we observed that the charge yield at a high electric field of $\sim
10^5~\textrm{Vcm}^2\textrm{g}^{-1}$ increases as the TMA fraction
increases, which we interpreted as the effect of Penning transfer. 
We also found that charge amplification starts at lower electric field
with the addition of TMA.
This could be due to lower ionization potential of TMA and also due to
Penning transfer.

\begin{figure}[h]
\hspace{3pc}\includegraphics[width=18.5pc]{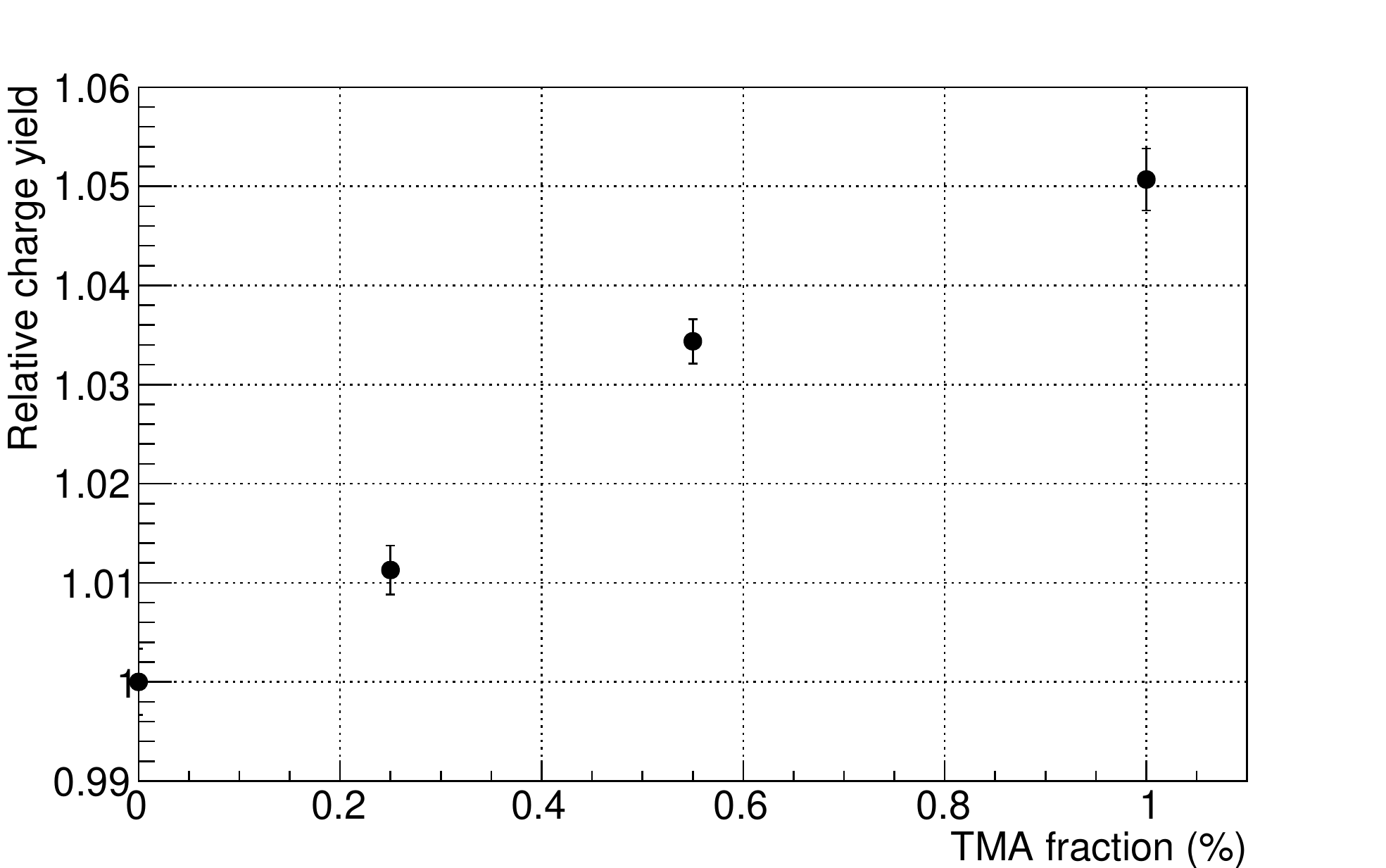}\hspace{1pc}%
\begin{minipage}[b]{14pc}
\caption{\label{fig:penning-tma}Relative size of averaged charge
  as a function of TMA concentration for a  total pressure of approximately 4 bar.
  The data for
  $10^5 < E/\rho < 2 \times 10^5~(\textrm{Vcm}^2\textrm{g}^{-1})$ are
  used to avoid the effect of the recombination process.}
\end{minipage}
\end{figure}

Figure~\ref{fig:penning-tma} shows the relative size of charge yield for
$10^5 < E/\rho < 2 \times 10^5~(\textrm{Vcm}^2\textrm{g}^{-1})$, as a
function of TMA concentration. We found a clear correlation between charge
increase and TMA concentration. The charge yield is increased by $\sim
5$\% with 1\% TMA concentration. 
In gaseous Xe, the relative size of excitation and ionization is measured
to be $1/W_{sc} : 1/W_{i} \sim 1: 2.5$~\cite{Renner:2014mha}, where
$W_{sc(i)}$ is the amount of energy loss of electrons needed to excite (ionize) one Xe
atom. Penning efficiency, $\epsilon (Penning)$, can be calculated
using the relative charge yield between Xe+TMA mixture to pure Xe,
$I(\textrm{Xe+TMA})/I(\textrm{pure Xe})$, as
\begin{equation}
  \label{eq:1}
  \epsilon (Penning) = \frac{W_{sc}}{W_i}
  \left(\frac{I(\textrm{Xe+TMA})}{I(\textrm{pure Xe})} -1\right).
\end{equation}
Hence, a 5\%
increase in the charge yield translates into 10-15\% of Penning transfer
efficiency. 
We observed a clear signature of Penning transfer. However, the amount of
Penning transfer is found to be 
relatively small to make any significant impact on the ionization signal.


\subsection{Light yield}

\begin{figure}[h]
\includegraphics[width=22pc]{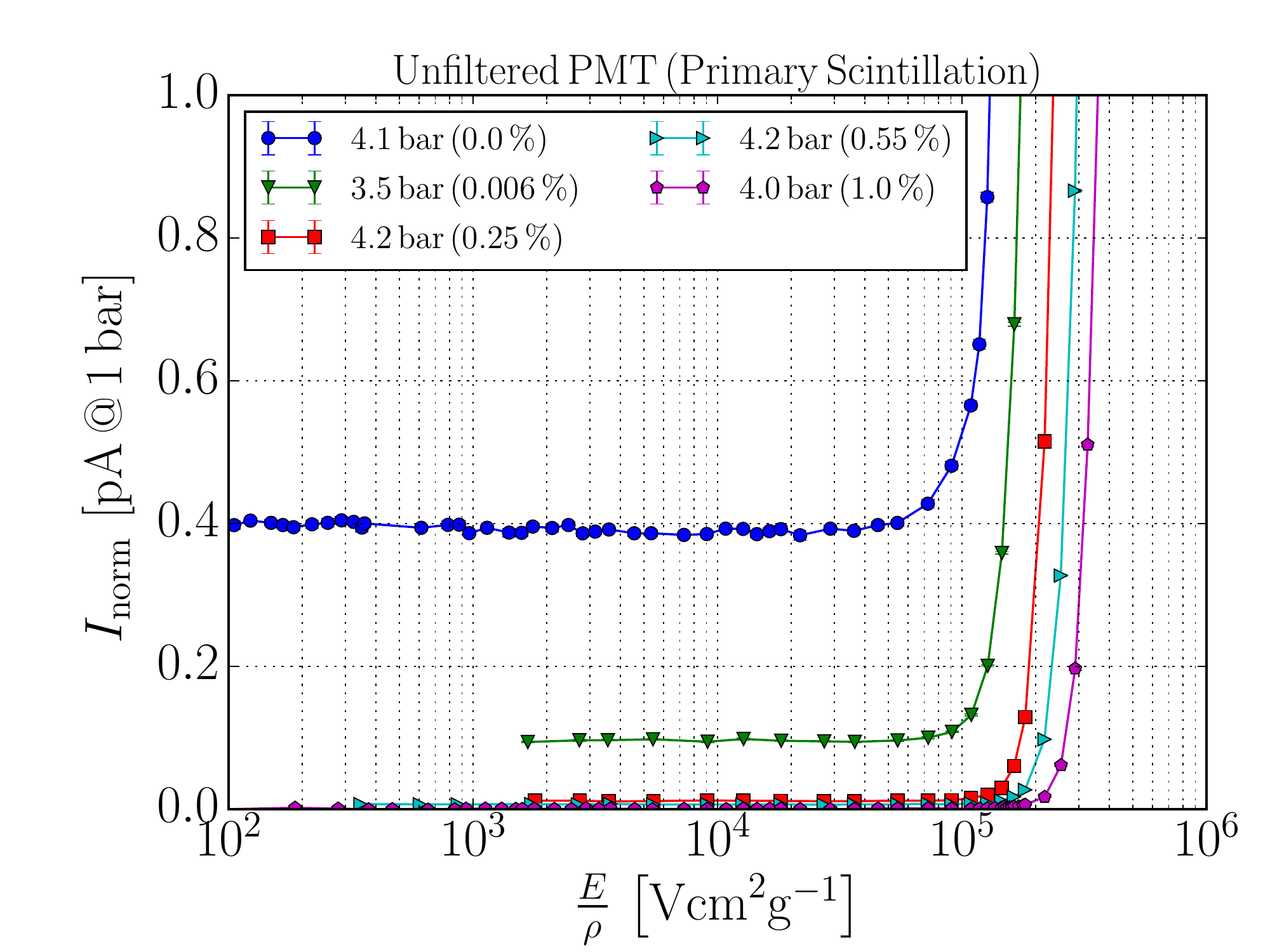}\hspace{1pc}%
\begin{minipage}[b]{14pc}\caption{\label{fig:XeTMA-4bar-S1}
    Primary scintillation light yield with Xe+TMA gas mixture,
    measured at approximately 4~bar total pressure and various TMA concentration.}
\end{minipage}
\end{figure}

\begin{figure}[h]
\begin{center}
\includegraphics[width=18pc]{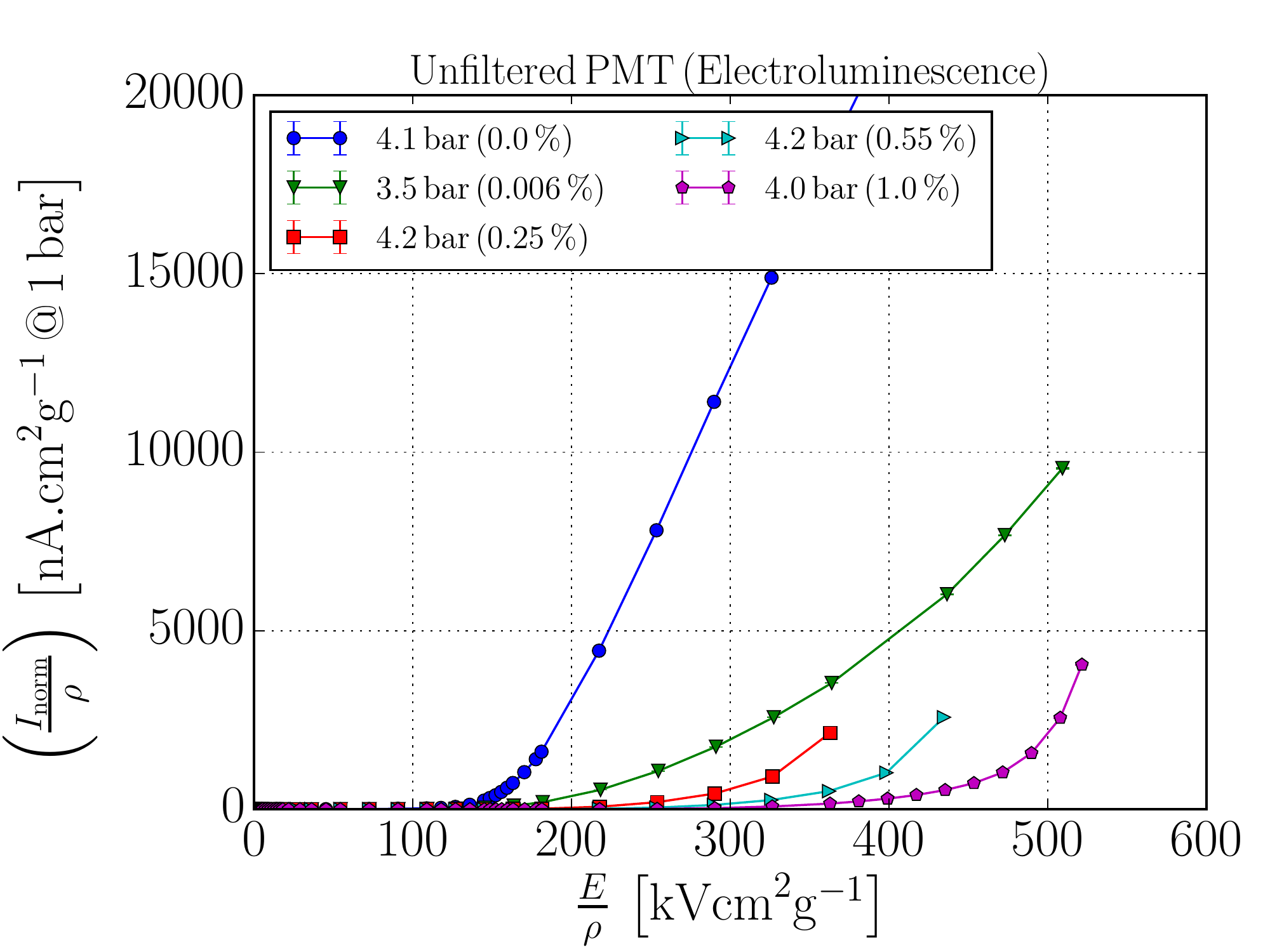}
\includegraphics[width=18pc]{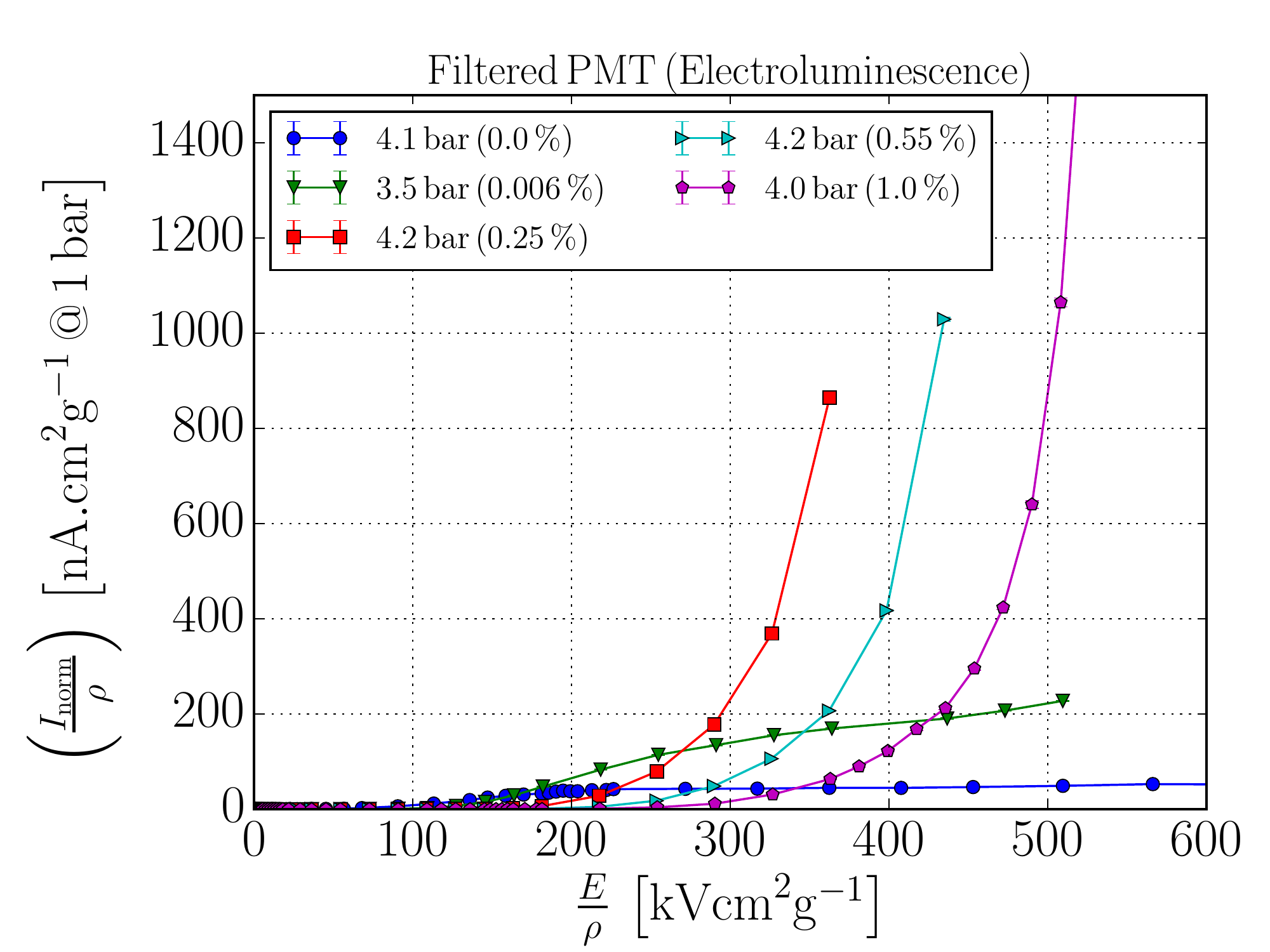}
\caption{\label{fig:XeTMA-4bar-EL} Electroluminescence light yield
  with Xe+TMA gas mixture,
  measured at approximately 4~bar total pressure and for various TMA
  concentrations. The left(right) panel shows the results from unfiltered(filtered) PMT.}
\end{center}
\end{figure}

Figure~\ref{fig:XeTMA-4bar-S1} shows results of light yield
measurements with various TMA concentrations with an unfiltered PMT. 
We found a large reduction of primary scintillation light once a very
small TMA fraction, as low as $\sim0.01\%$, is introduced. 
On the other hand, we observed electroluminescence light at
higher electric field, as shown in Fig.~\ref{fig:XeTMA-4bar-EL}.
This EL light is observed for both unfiltered and filtered PMTs, which
indicates that it is scintillation light from TMA and the light from
Xe is practically lost.

\begin{figure}[h]
\includegraphics[width=23pc]{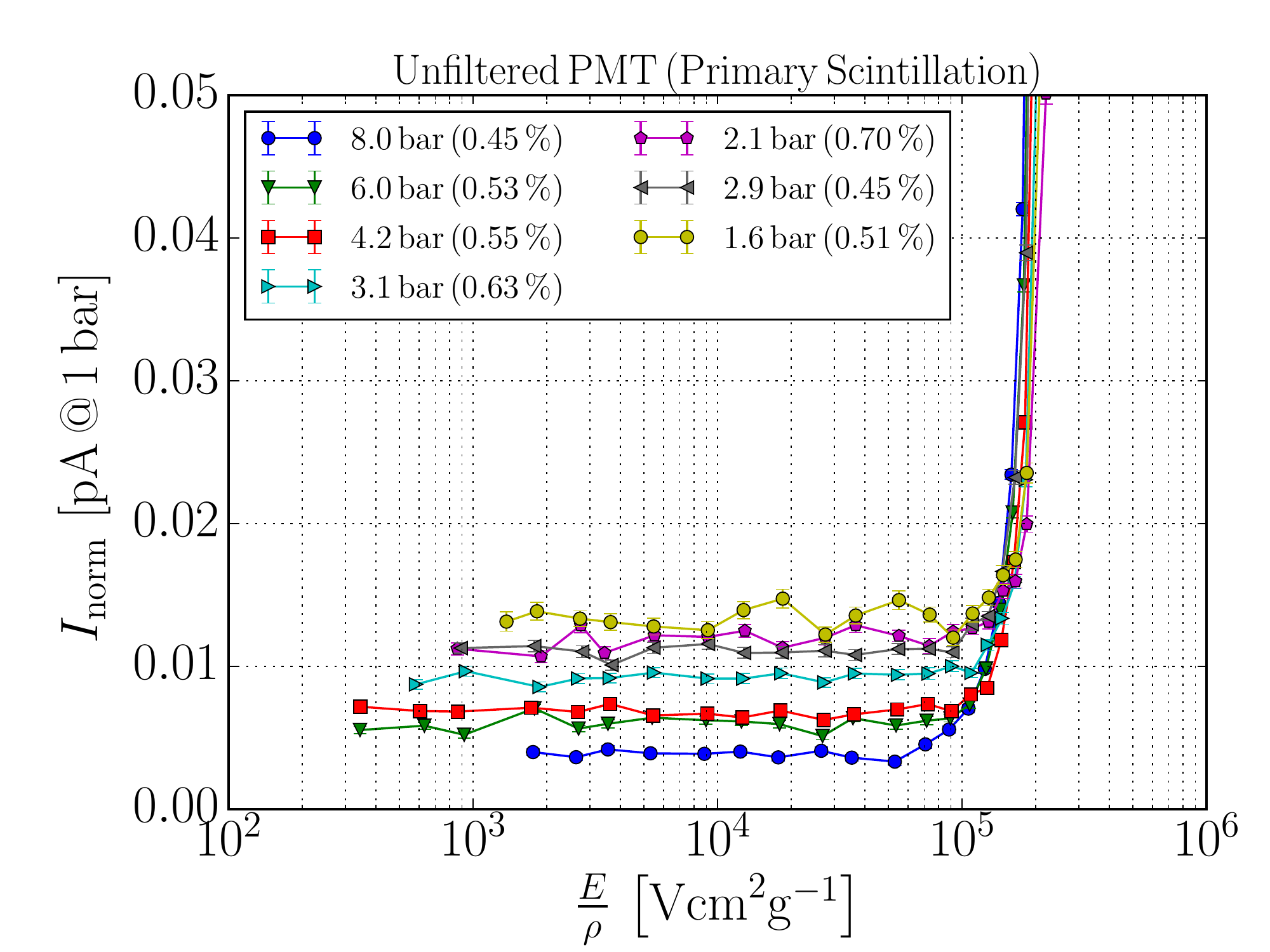}\hspace{1pc}%
\begin{minipage}[b]{14pc}\caption{\label{fig:XeTMA-S1}
    Primary scintillation light yield with Xe+TMA gas mixture. 
  Results of unfiltered PMT at approximately 0.5\% TMA concentration
  and for different total pressures are shown.}
\end{minipage}
\end{figure}

Although the amount of primary scintillation light is largely suppressed
with Xe+TMA mixture, we observed non-zero amount of the
scintillation light. 
Figure~\ref{fig:XeTMA-S1} shows the observed current from the unfiltered
PMT with $\sim 0.5\%$ TMA concentration for different total pressures. 
The size of the signal is approximately 1-4\% of that of pure Xe. This is
roughly consistent with the expectation from direct excitation of TMA,
after taking into account the difference of PMT quantum efficiency between 170~nm
and 300~nm and the difference in excitation cross sections between Xe and TMA.
A consistent amount of scintillation light was observed with the filtered
PMT as well, which also supports the conclusion that the observed
light is only emitted from TMA.
The suppression of the light yield at higher pressure is consistent with
the self-quenching effect of the TMA~\cite{TMA-selfquench}. A more careful
investigation of this self-quenching effect is underway.
If the TMA recombination produces fluorescence light, it would
be anti-correlated to the charge yield and would appear as a negative slope
of light yield as a function of electric field.
However, we
observed a constant light yield as a function of electric field
for $10^3 < E/\rho < 10^5~(\textrm{Vcm}^2\textrm{g}^{-1})$.
Therefore, we conclude that no significant additional scintillation light, such as light from TMA
recombination, energy transfer from Xe, is observed.

\section{Discussion}
One of the most significant effects that we observed with the Xe and
TMA mixture is the high suppression of the primary scintillation
light. It is likely that TMA is either absorbing Xe scintillation
light, or taking out energy from Xe excited states but not releasing
the energy in the form of observable light.
We observed enhancement of the recombination process through the reduction
of charge yield at low electric field. However, we did not observe
a corresponding increase of scintillation light. This can be due to
either lack of charge-exchange process ($\textrm{Xe}^+ +
\textrm{TMA}\to \textrm{Xe} + \textrm{TMA}^+$) and also due to 
lack of fluorescence after TMA recombination.
An increase of charge yield due to Penning effect  was
clearly observed, and Penning efficiency was found to be 10-15\% for 1\%
TMA concentration at 4~bar total pressure. This, however, corresponds
to an increase in the number of ionization electrons
at the level of only 5\% and would not
give significant improvement in the detector performance. 
All of those observations indicate that, unfortunately, TMA is not a suitable additive
to enhance performance of gaseous Xe TPC for dark matter
searches.
While the suppression of scintillation light is also problematic for
neutrinoless double beta decay searches, 
other features of TMA, such as reduction of diffusion and
increase of charge yield, can still be beneficial for this application
and are actively studied~\cite{Gonzalez-Diaz:2015oba,Alvarez:2013kqa}.

On the other hand, our system is capable of measuring scintillation
and ionization yield for any gas mixture at high precision. Other gas
mixtures, such as Xe and Triethylamine(TEA), may be tested to explore ideal gas mixtures.

\section{Summary}

Using a test parallel-plate ionization chamber, we successfully
made high-precision measurement of scintillation and ionization yield for
Xe+TMA gas mixture at various electric fields.
Although we found 
the Penning effect and enhancement of the recombination process
with the addition of TMA, many undesired features, such as  
strong suppression of the scintillation light and no sign of recombination
light, were also found. 
Therefore, it is likely difficult to utilize TMA to enhance the performance
of gaseous Xe TPCs in dark matter searches. Other features of TMA, such as
diffusion reduction and charge gain increase, can still benefit
neutrinoless double beta decay searches.

Although TMA is found to be not an ideal additive, our system has
proven to be capable of measuring 
scintillation and ionization yield at high precision. We are exploring
the possibility of testing other gas mixtures, such as Xe+TEA.

\section{Acknowledgments}
We acknowledge the indispensable technical support from Tom Miller (LBNL) during the design, construction and running of the detector. This work was supported by the Director, Office of Science, Office of Basic Energy Sciences, of the US Department of Energy under contract no. DE-AC02-05CH11231. J. Renner (LBNL) acknowledges the support of a US DOE NNSA Stewardship Science Graduate Fellowship under contract no. DE-FC52-08NA28752.

\section*{References}

\bibliographystyle{iopart-num}
\bibliography{nakajima_TPC2014}

\end{document}